# Ballistic gelatin as a putative substrate for EEG phantom devices

W. David Hairston, Geoffrey A. Slipher, and Alfred B. Yu

*Abstract*—Phantom devices allow the human variable to be controlled for in order to allow clear comparison and validation of biomedical imaging hardware and software. There is currently no standard phantom for electroencephalography (EEG). To be useful, such a device would need to: (a) accurately recreate the real and imaginary components of scalp electrical impedance, (b) contain internal emitters to create electrical dipoles, and (c) be easily replicable across various labs and research groups. Cost-effective materials, which are conductive, repeatable, and easily formed are a missing key enabler for EEG phantoms. Here, we explore the use of ballistics gelatin, an inexpensive, easily-formable and repeatable material, as a putative substrate by examining its electrical properties and physical stability over time. We show that varied concentrations of NaCl salt relative to gelatin powder shifts the phase/frequency response profile, allowing for selective "tuning" of the material electrical properties.

## I. Background

The development of several new approaches to EEG data acquisition, especially for "real-world neuroimaging"[1], has emphasized the need for a standardized "phantom" for use in EEG. We propose the use of ballistic gelatin (BG) as a conductive substrate for such a device. While BG is primarily used for its mechanical properties resembling that of living tissue [2], [3], it is also inherently electrically conductive [2], [4]. This feature, in addition to being relatively inexpensive and easy to form suggests that it may be a suitable medium. Little attention has been paid to the degree to which ballistics gelatin can be selectively "tuned" to match the electrical properties of various tissues. We present data showing how BG electrical properties are tunable using NaCl concentration.

## II. Methods

NaCl was added at varied concentrations to water heated to ~100° C and mixed with ballistic gelatin powder at a ratio of 10/1 water/powder by mass. Samples were set in molds yielding 20mm thick samples with smooth, flat contact surfaces. Impedance magnitude and phase angle information were collected for the disc-shaped gelatin samples using a Keyence E5061B-LF Network analyzer. The samples were interrogated with the network analyzer between 5Hz and 1500Hz using a parallel plate electrode configuration with the gelatin samples sandwiched in the middle. Figure 1 presents results for impedance magnitude as a function of frequency for 6 NaCl concentrations across the typical spectra of EEG. Here we see a dramatic (>100x) reduction in impedance magnitude with increased NaCl, suggesting significant tunability. The response is also non-linear across frequency, similar to biological tissue [5]. Significant phase shifts as well as non-linear trends in phase are also observed. The similarity to biological tissue and tunability allows a wide range of scalp conditions to be simulated on a phantom EEG device in an inexpensive and repeatable manner. One potential challenge for BG as a substrate is its durability over time. Measurement of mass loss over the course of 384 hours suggests an average rate of roughly 0.3% per day kept refrigerated and covered when not in use.

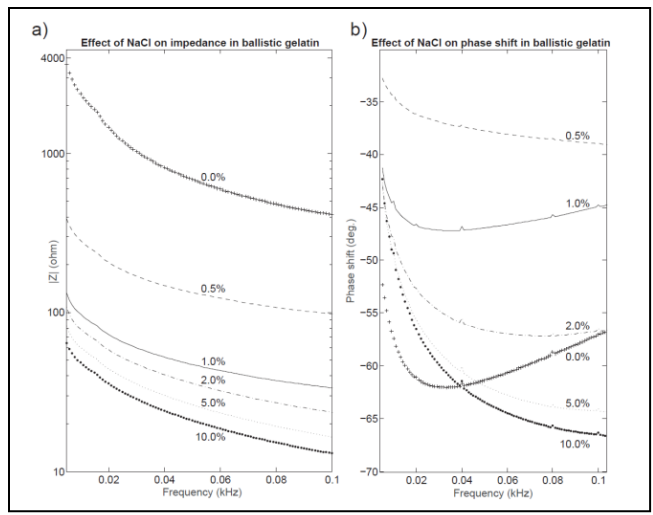

Figure 1. Effect of NaCl loading on ballistic gelatin impedance magnitude (a), and phase (b).

A full-size EEG phantom was created using a 3-part mold 3-D printed based on the MRI of a human volunteer (design files are available upon request). With a printed mold on hand, fabrication takes 3 hours plus 12 hours cooling time, and costs only US$20 in material.

## III. Summary

This work demonstrates the feasibility of using BG as a cost-effective surrogate conductive material for creating EEG phantom devices. Through the addition of NaCl loading as an ion source, the material can be tuned to broadly match a desired impedance profile, then easily formed into a desired shape with embedded wire conductors.

W. D. Hairston and A. B. Yu. (corresponding), are with the Human Research and Engineering Directorate, U.S. Army Research Laboratory, Aberdeen Proving Ground, MD 21005 (Ph: 410-278-5925; email: alfred.b.yu.civ@mail.mil).

G. A. Slipher is with the Vehicle Technology Directorate, U.S. Army Research Laboratory, APG, MD 21005;